\documentclass[letterpaper,11pt]{article}

\usepackage{amssymb}
\usepackage{amsmath}
\usepackage{cite}

\numberwithin{equation}{section}

\newcommand {\beq} {\begin{equation}}
\newcommand {\eeq} {\end{equation}}
\newcommand {\beqa} {\begin{eqnarray}}
\newcommand {\eeqa} {\end{eqnarray}}
\newcommand {\beqan} {\begin{eqnarray*}}
\newcommand {\eeqan} {\end{eqnarray*}}

\newcommand {\nn} {\nonumber}
\newcommand {\ph}[1]{\phantom{#1}}
\newcommand {\ie}{i.e.~}
\newcommand {\eg}{e.g.~}

\newcommand {\sss} {\scriptscriptstyle}

\newcommand{\ordo}{\ensuremath{\mathcal{O}}}

\newcommand{\pbf}{\ensuremath{\mathbf{p}}}
\newcommand{\nbf}{\ensuremath{\mathbf{n}}}

\newcommand{\ah}{\ensuremath{\hat{a}}}
\newcommand{\bh}{\ensuremath{\hat{b}}}
\newcommand{\ch}{\ensuremath{\hat{c}}}
\newcommand{\dha}{\ensuremath{\hat{d}}}
\newcommand{\eh}{\ensuremath{\hat{e}}}

\newcommand{\Xh}{\ensuremath{\hat{X}}}

\newcommand{\al}{\ensuremath{\alpha}}
\newcommand{\be}{\ensuremath{\beta}}
\newcommand{\ga}{\ensuremath{\gamma}}
\newcommand{\Ga}{\ensuremath{\Gamma}}
\newcommand{\de}{\ensuremath{\delta}}

\newcommand{\eps}{\ensuremath{\epsilon}}

\newcommand{\ka}{\ensuremath{\kappa}}

\newcommand{\si}{\ensuremath{\sigma}}
\newcommand{\Si}{\ensuremath{\Sigma}}

\newcommand{\Om}{\ensuremath{\Omega}}
\newcommand{\Th}{\ensuremath{\Theta}}

\newcommand{\alh}{\ensuremath{\hat{\alpha}}}
\newcommand{\beh}{\ensuremath{\hat{\beta}}}
\newcommand{\gah}{\ensuremath{\hat{\gamma}}}
\newcommand{\deh}{\ensuremath{\hat{\delta}}}
\newcommand{\epsh}{\ensuremath{\hat{\epsilon}}}
\newcommand{\Thh}{\ensuremath{\hat{\Theta}}}

\newcommand{\ald}{\ensuremath{\dot{\alpha}}}
\newcommand{\bed}{\ensuremath{\dot{\beta}}}
\newcommand{\gad}{\ensuremath{\dot{\gamma}}}
\newcommand{\ded}{\ensuremath{\dot{\delta}}}
\newcommand{\ad}{\ensuremath{\dot{a}}}
\newcommand{\bd}{\ensuremath{\dot{b}}}
\newcommand{\cd}{\ensuremath{\dot{c}}}
\newcommand{\dd}{\ensuremath{\dot{d}}}

\newcommand{\mathHb}[1]{{\mathop{\kern0pt#1}\limits^{\,\sss
      \prime\prime}\vphantom{#1}}}

\newcommand {\we} {\wedge}
\newcommand {\pa} {\partial}

\newcommand{\eqnlab}[1]{\label{eqn:#1}}

\newcommand{\eqnref}[1]{(\ref{eqn:#1})}

\newcommand{\Eqnref}[1]{Eq.~(\ref{eqn:#1})}

\newcommand{\Eqsref}[1]{Eqs.~(\ref{eqn:#1})}

\newcommand{\seclab}[1]{\label{sec:#1}}

\newcommand{\secref}[1]{\ref{sec:#1}}

\begin{document}

 \pagestyle{empty}
 \vskip-10pt

\begin{center}

\vspace*{2cm}

\noindent
{\LARGE\textsf{\textbf{Scattering amplitudes for particles and strings
      \\[5mm] in six-dimensional (2,0) theory}}}
\vskip 1truecm
\vskip 2truecm

{\large \textsf{\textbf{Erik Flink}}} \\ 
\vskip 1truecm
{\tt erik.flink@chalmers.se}
\vskip 0.5truecm
{\it Department of Fundamental Physics\\ Chalmers University of
  Technology\\ SE-412 96 G\"{o}teborg,
  Sweden}\\
\end{center}
\vskip 1cm
\noindent{\bf Abstract:}
We study the scattering of low-energy tensor multiplet particles against a BPS saturated cosmic string. We show that the corresponding $S$-matrix is largely determined by symmetry considerations. We then apply a specific supersymmetric model of (2,0) theory and calculate the scattering amplitudes to lowest non-trivial order in perturbation theory. Our results are valid as long as the energy of the incoming particle is much lower than the square root of the string tension. The calculation involves the quantization of a (2,0) tensor multiplet and the derivation of an effective action describing the low-energy particles in the presence of a nearly BPS saturated string.

\newpage
\pagestyle{plain}

\tableofcontents

\section{Introduction}

The six-dimensional (2,0) theories \cite{Witten:1995} offer a very interesting arena to study strings and quantum field theory in higher dimensions, without having to take the effects of dynamical gravity into account. In spite of this simplification, the theories are quite difficult to approach, especially in the most interesting superconformal fixed point where the strings are tensionless: We don't even know what a tensionless string really is. If we instead restrict our attention to a generic point in the moduli space, where the conformal symmetry is broken, the (2,0) superalgebra has a realization in both a massless tensor particle multiplet and a tensile string vector multiplet \cite{Gustavsson:2001b}. This indicates that the degrees of freedom should consist of both massless particles and strings, the tension of which is given by the norm of the vacuum expectation values of the scalar moduli fields. Furthermore, the tensor multiplet contains a two-form gauge field with self-dual three-form field strength, which couples naturally to strings and not to particles. Supersymmetry then suggests that none of the fields in the tensor multiplets couple directly to each other. Instead, all particle interactions are mediated by the strings. A final intricate property of these theories follows by combining the self-duality of the three-form field strength with Dirac quantization effects: The 'electromagnetic' coupling constant is of order unity. This is a substantial problem, since it seems to render a perturbative approach impossible.

Our research is based on the fact that an infinitely long tensile string (propagating in six-dimensional Minkowski spacetime) has infinite energy, and therefore cannot be pair-created in any processes where the involved energies $E$ are finite. Hence, the Hilbert space is divided into different sectors characterized by the kinds of strings they contain. Each of these sectors constitute a separate unitary quantum theory. By further restricting the energy $E$ to be much smaller than the square root of the string tension $T$, neither closed strings can be pair-created in any scattering events. This is because, on dimensional grounds, the energy of a closed string must be of the order $\sqrt{T}$. We may therefore neglect the influence of closed strings in a perturbative approach, using the dimensionless quantity $E/\sqrt{T}$ as the expansion parameter. The different sectors of the perturbative part of the Hilbert space are then characterized by the number of infinitely long strings they contain. The ground state of each sector of the Hilbert space consists of BPS saturated strings, meaning that they are exactly straight. The possible excitations of the ground states are two-fold, namely string waves propagating on the world-sheet and tensor multiplet particles. In the limit where $T$ is infinite, the string waves and particles decouple, and we are left with the well understood theory of a free tensor particle multiplet and a non-trivial two-dimensional conformal theory of string waves \cite{Henningson:2004}.

To further investigate the perturbative part of the Hilbert space, it is natural to consider the $S$-matrix for scattering of low-energy particles and tensile strings. In this paper we content ourselves with studying processes in which the in-state consists of a single particle and a BPS saturated string in the infinite past, and the out-state consists of another particle and the same kind of BPS saturated string, in the infinite future. We show in section \secref{kinematics} that the $S$-matrix for such scattering processes is highly constrained by symmetry considerations. In fact, it is determined up to a single arbitrary function that depends only on the dimensionless ratio $E^2/T$ together with the angle between the momentum of the incoming particle and the string and the angle between the momenta of the incoming and outgoing particles. To determine this function we must make use of a specific model, though. The interaction terms constructed in \cite{AFH:2003sc} are precisely what we need. They describe a string coupled to an on-shell tensor multiplet background. By integrating out the dependence on the string waves, we obtain an effective action describing small fluctuations of the background in the presence of a nearly straight infinitely long string. It is then fairly straight forward to obtain the desired $S$-matrix elements from this effective action. These calculations show perfect agreement with the symmetry analysis. Furthermore, they make evident that the scattering process of a single tensor multiplet particle against a string takes two fundamentally different ways: Some particles are first converted into off-shell string waves, which live for a short while and then reconvert into other tensor multiplet particles. Other particles simply bounce off the string without exciting any string waves. Finally, we note that the calculated cross section for this kind of scattering is remarkably simple to lowest non-trivial order.

\section{Kinematics of particle-string scattering}
\seclab{kinematics}
As described in the introduction, we will consider the amplitudes for scattering a single tensor multiplet particle off a string in one of its ground states. In this section, we will determine the restrictions on such amplitudes that follow from the symmetry of the problem. To begin with, we will focus our attention on the bosonic subalgebra of the $(2, 0)$ superconformal algebra. This is isomorphic to $so (6, 2) \times so (5)_R$, where the first factor is the conformal algebra in six dimensions, and the second factor is the $R$-symmetry algebra. This symmetry is partly spontaneously broken, though, as we will now review.

First of all, a (2,0) tensor multiplet consists of five scalar fields $\phi$ transforming as a vector under the $SO(5)$ $R$-symmetry group. It also contains a two-form gauge field $b$ with self-dual three-form field strength $h$, both of which are singlets under the $SO(5)$ $R$-symmetry. Upon quantization, these fields yield eight bosonic degrees of freedom, and supersymmetry requires that there should be equally many fermionic ones. These are realized as a symplectic Majorana spinor $\psi$, which is a Weyl spinor under $SO(5,1)$ and a spinor also of $SO(5)_R$.

Furthermore, the values of the scalar fields at spatial infinity, \ie the vacuum expectation values, constitute the moduli of (2,0) theory. Let us denote the direction of this $SO(5)_R$ vector by $\phi_\infty$ and its magnitude by $T$. It then follows that a non-vanishing value of $T$ breaks the conformal group to the six-dimensional Poincar\'e group ${\mathbb R}^{5, 1} \times SO (5, 1)$. It also breaks the $R$-symmetry group to a subgroup $SO (4)_R$.  In a situation with such non-zero moduli, we may also introduce tensile strings. The presence of a straight static string along the spatial direction given by the unit vector ${\bf n}$ breaks the Poincar\*e group further to an ${\mathbb R}^{1, 1} \times SO (1, 1) \times SO (4)_{\bf n}$ subgroup, where the first two factors constitute the world-sheet Poincar\'e group, and the last factor consists of spatial rotations in the directions transverse to ${\bf n}$. (There is  in fact a degenerate multiplet of various polarization states of straight static strings \cite{Gustavsson:2001b}, but this degeneracy will not play any role in the present paper. The string polarization does not change in a particle scattering process, and does not affect the scattering amplitudes.)

Next, we consider a massless tensor multiplet particle of momentum ${\bf p}$. (We now disregard the symmetry breaking by moduli fields and strings discussed in the previous paragraph.) This breaks the Lorentz group to the group $SO (4)_{\bf p}$ of spatial rotations in the directions transverse to ${\bf p}$. There is a degenerate multiplet of orthonormal particle states $\left| {\bf p}, s \right>$, where the polarization label $s$ transforms under $SO (4)_{\bf p} \times SO (5)_R$ as
\beq
\eqnlab{tm polarizations}
(1, 5) \oplus (2, 4) \oplus (3_+, 1) .
\eeq
(The scalar, chiral spinor, anti-chiral spinor, self-dual tensor, anti self-dual tensor, and vector representations of $SO (4)$ are denoted as $1$, $2$, $2^\prime$, $3_+$, $3_-$, and $4$. The scalar, spinor, and vector representations of $SO (5)$ are denoted as $1$, $4$, and $5$.) In a field theory description of these polarizations, the three terms in \Eqnref{tm polarizations} correspond to the fields $\phi$, $\psi$ and $b$, respectively.

We need to describe more explicitly how various symmetry transformations act on the particle states $\left| {\bf p}, s \right>$. By a scale transformation, we can restrict to the case where the momentum ${\bf p}$ is a unit vector. In the presence of a string along the direction ${\bf n}$, we may define a symmetry transformation $L ({\bf p})$ as the spatial rotation in the plane spanned by ${\bf n}$ and ${\bf p}$ by the angle $\theta$ between ${\bf n}$ and ${\bf p}$. We now define the particle state $\left| {\bf p}, s \right>$ in terms of the particle state $\left| {\bf n}, s \right>$ as
\beq
\left| {\bf p}, s \right> = L ({\bf p}) \left| {\bf n}, s \right> .
\eeq
Let now $\Lambda$ be an element of $SO (4)_{\bf n}$, \ie the group of spatial rotations in the directions transverse to  ${\bf n}$. It acts on a state of the form $\left| {\bf n}, s \right>$ by a transformation of the $s$ quantum number:
\beq
\Lambda \left| {\bf n}, s \right> = \left| {\bf n},  \Lambda s \right> .
\eeq
Its action on an arbitrary state $\left| {\bf p}, s \right>$ can now be computed as
\beq
\Lambda \left| {\bf p}, s \right> = \Lambda L ({\bf p}) \left| {\bf n}, s \right> = L (\Lambda {\bf p}) \Lambda \left| {\bf n}, s \right>  = L (\Lambda {\bf p})  \left| {\bf n},  \Lambda s \right> = \left| \Lambda {\bf p},  \Lambda s \right> ,
\eeq
where we have used that $L^{-1}(\Lambda \pbf) \Lambda L(\pbf) = \Lambda$, in the second step.

To construct a basis for the polarization label $s$, it is convenient to use, in addition to the string direction ${\bf n}$ and the momentum ${\bf p}$ of e.g. an incoming particle, also the momentum ${\bf p}^\prime$ of an outgoing particle. We then define $\hat{J}$ as the generator of the group $SO (2)_{\bf n, p , p^\prime}$ of spatial rotations in the plane orthogonal to ${\bf n}$, ${\bf p}$, and ${\bf p}^\prime$ normalized so that $\exp i \alpha \hat{J}$ is a rotation by an angle $\alpha$. The particle polarization $s$ can now be characterized by giving the representation of $SO (4)_R$, with the $\hat{J}$ eigenvalue as a subscript. The only ambiguity is that there are two $SO (4)_R$ singlet states with $\hat{J}$ eigenvalue zero. To distinguish between them, we denote the states that originate from the $(3_+, 1)$ representation of $SO (4)_{\bf p} \times SO (5)_R$ as $1^\prime_{+1}$, $1^\prime_{0}$, and $1^\prime_{-1}$, whereas the states that originate from the $(1, 5)$ representation are  denoted as $1_0$ and $4_0$. As an orthonormal basis for the states $\left| {\bf p}, s \right>$ , we can then use
\beq
\begin{array}{ccccc}
 & & \left| {\bf p}, 1^\prime_{+1} \right> & & \cr
 &  \left| {\bf p}, 2^\prime_{+\frac{1}{2}} \right> & &  \left| {\bf p}, 2_{+\frac{1}{2}} \right> & \cr
  \frac{1}{\sqrt{2}} (\left| {\bf p}, 1_0 \right> -  \left| {\bf p}, 1^\prime_0 \right> ) & &  \left| {\bf p}, 4_0 \right> & & \frac{1}{\sqrt{2}} (\left| {\bf p}, 1_0 \right> +  \left| {\bf p}, 1^\prime_0 \right>) \cr
 &  \left| {\bf p}, 2_{-\frac{1}{2}} \right> & &  \left| {\bf p}, 2^\prime_{-\frac{1}{2}}  \right> & \cr
 & &  \left| {\bf p}, 1^\prime_{-1} \right>
 \end{array} . \label{sbasis}
 \eeq
 
 Our aim is now to compute the $S$-matrix elements ${}_{\rm in} \left< {\bf p}, s \right| \left. {\bf p^\prime}, s^\prime \right>_{\rm out}$ that describe how a tensor multiplet particle is scattered off a string along the direction ${\bf n}$. We define $\theta$ as the angle of incidence, \ie the angle between ${\bf n}$ and ${\bf p}$. We also define $\rho$ through the relation
\beq
\cos \varphi = \sin^2 \theta \cos \rho + \cos^2 \theta ,
\eeq 
where $\varphi$ is the scattering angle, \ie the angle between ${\bf p}$ and ${\bf p}^\prime$. Scaling properties, conservation of momentum along ${\bf n}$, conservation of energy, and rotational invariance (almost) implies that the probability amplitude for an in-state $\left| {\pbf}, s \right>_{\rm in}$ to become an out-state $\left| {\pbf'}, s' \right>_{\rm out}$ must be of the form
\beq
\eqnlab{prob}
\mathcal{P}(\pbf,s \rightarrow \pbf',s') = |\pbf|^2 \delta (|{\bf p}| - |{\bf p}^\prime|) \delta ({\bf p} \cdot {\bf n} - {\bf p}^\prime \cdot {\bf n}) \delta_{s s^\prime} \tilde{f} (|{\bf p}|^2 / T, \theta, \rho, s) 
\eeq
for some function $\tilde{f}$. The only exception appears to be a possible transition between the states $\frac{1}{\sqrt{2}} (\left| {\bf p}, 1_0 \right> -  \left| {\bf p}, 1^\prime_0 \right>)$ and $\frac{1}{\sqrt{2}} (\left| {\bf p}^\prime, 1_0 \right> +  \left| {\bf p}^\prime, 1^\prime_0 \right>)$, or between the states $\frac{1}{\sqrt{2}} (\left| {\bf p}, 1_0 \right> +  \left| {\bf p}, 1^\prime_0 \right>)$ and $\frac{1}{\sqrt{2}} (\left| {\bf p}^\prime, 1_0 \right> -  \left| {\bf p}^\prime, 1^\prime_0 \right>)$. As we will see below, this does not occur. This is the reason for our choice of basis for the polarization label $s$.

We now focus on the $S$-matrix, the elements of which we write as
\beq
S_{\pbf',s':\pbf,s} = {}_{\rm out}\left< \pbf',s' \right| \left. \pbf,s \right>_{\rm in} =  \de^{(5)}(\pbf-\pbf') \de_{ss'} + i T_{\pbf',s':\pbf,s}.
\eeq
From the above, it then follows that the $T$-matrix elements must be of the form
\beq
\eqnlab{t matrix form}
T_{\pbf',s':\pbf,s} = \frac{1}{(2\pi|\pbf|)^3} \delta (|{\bf p}| - |{\bf p}^\prime|) \delta ({\bf p} \cdot {\bf n} - {\bf p}^\prime \cdot {\bf n}) \mathcal{M}_{ss'}(|{\bf p}|^2 / T, \theta, \rho),
\eeq
where the factor $|\pbf|^{-3}$ is needed to get the dimensions right. According to \Eqnref{prob}, the matrix $\mathcal{M}$ is diagonal (for the basis in (\ref{sbasis})). Its elements are commonly referred to as the invariant matrix elements. Further constraints on $\mathcal{M}$ follow from considering fermionic symmetries. The fermionic generators of the $(2, 0)$ superconformal algebra transform as a spinor under the $R$-symmetry group $SO (5)_R$. They transform as a chiral spinor under the $SO (6, 2)$ conformal group, which amounts to a Dirac spinor under the $SO (5, 1)$ Lorentz group, but we will only be interested in the part which transforms as a chiral spinor under the latter group, containing the generators of supersymmetry.\footnote{The part corresponding to the anti-chiral spinor contains the generators of special supersymmetry.} With respect to the subgroup $SO (4)_{\bf n} \times SO (4)_R$, we find that the fermionic generators transform in the representation
\beq
(2, 2) \oplus (2^\prime, 2^\prime) \oplus (2, 2^\prime) \oplus (2^\prime, 2) . \label{Qrep}
\eeq

We will first consider the action of such a fermionic generator $Q^0$ on a particle state of the form $\left| {\bf n}, s \right>$, describing a particle propagating along the string direction. We denote this as
\beq
Q^0 \left| {\bf n}, s \right> = \sum_{s^\prime} (Q^0)_{s s^\prime} \left| {\bf n}, s^\prime \right> ,
\eeq
where $(Q^0)_{s s^\prime}$ are the matrix elements of the operator $Q^0$ relative to our basis. For $Q^0$ in the representation $(2^\prime, 2)$ or $(2^\prime, 2^\prime)$, these are in fact all zero, \ie the BPS-saturated states $\left| {\bf n}, s \right>$ are annihilated by such a generator. For  $Q^0$ in the $(2, 2)$ and $(2, 2^\prime)$ representions, we may label the generators by the representation of $SO (4)_R$ with the $\hat{J}$ eigenvalue as a subscript and arrange them as follows:
\beq
\begin{array}{cc}
2_{+ \frac{1}{2}} \nwarrow &  \nearrow 2^\prime_{+ \frac{1}{2}} \cr
2^\prime_{- \frac{1}{2}} \swarrow & \searrow 2_{- \frac{1}{2}} 
\end{array}
\eeq
The arrows indicate the direction of the action of the generator if the states $\left| {\bf n}, s \right>$ are arrranged as described above. All non-vanishing matrix elements  of $Q^0$ are given by $(Q^0)_{s s^\prime} = \sqrt{|{\bf p}|} = 1$, so that the supersymmetry algebra $\{Q, Q \} = P$ is fulfilled. 

The action of a fermionic generator $Q^0$ on an arbitrary state $\left| {\bf p}, s \right>$ can now be computed as
\beq
Q^0 \left| {\bf p}, s \right> = Q^0 L ({\bf p}) \left| {\bf n}, s \right> = L ({\bf p}) Q \left| {\bf n}, s \right> = L ({\bf p}) \sum_{s^\prime} (Q)_{s s^\prime} \left| {\bf n}, s^\prime \right>  =  \sum_{s^\prime} (Q)_{s s^\prime} \left| {\bf p}, s^\prime \right>  ,
\eeq
where the fermionic generator $Q$ is given by
\beq
Q = L^{-1} ({\bf p}) Q^0 L ({\bf p}) ,
\eeq
\ie $Q$ is obtained by acting with the spatial rotation $L ({\bf p})$ on $Q^0$. Obviously, we also have that
\beq
Q^0 \left| {\bf p}^\prime, s \right> = \sum_{s^\prime} (Q^\prime)_{s s^\prime} \left| {\bf p}^\prime, s^\prime \right> ,
\eeq
where 
\beq
Q^\prime = L^{-1} ({\bf p}^\prime) Q^0 L ({\bf p}^\prime) .
\eeq
Note that, when acting on a spinor,
\beqa
L ({\bf p}) & = & \exp (i \theta J) = \cos \frac{\theta}{2} + 2 i  J \sin  \frac{\theta}{2}  \cr
\eqnlab{spinor transf}
L ({\bf p}^\prime) & = & \exp (i \theta J^\prime) = \cos \frac{\theta}{2} + 2 i  J^\prime \sin  \frac{\theta}{2} 
\eeqa
where $J$ and $J^\prime$ are the generators of rotations in the plane spanned by ${\bf n}$ and ${\bf p}$ and ${\bf n}$ and ${\bf p}^\prime$ respectively. They are normalized so that $\exp i \alpha J$ or $\exp i \alpha J^\prime$ is a rotation by the angle $\alpha$. In the spinor representation, this means that  $J^2 = J^{\prime 2} = \frac{1}{4}$. When acting on a spinor with a definite chirality in the directions transverse to ${\bf n}$, the first term in $L ({\bf p})$ or $L ({\bf p}^\prime)$ preserves the chirality, whereas the second term reverses it.

We will only consider generators $Q^0$ in the representation
\beq
(2, 2^\prime) \oplus (2^\prime, 2) .
\eeq
that are unbroken by the presence of the BPS-saturated string. Given such a fermionic generator $Q^0$, we now wish to find the relationship between the corresponding rotated generators $Q$ and $Q^\prime$ described above. In fact, since we are only interested in the action of these generators on particle states, which are annihilated by generators in the $(2^\prime, 2)$ and $(2^\prime, 2^\prime)$ representations, it is sufficient to consider the equivalence classes $[Q]$ and $[Q^\prime]$ of $Q$ and $Q^\prime$ modulo these parts. 

Consider first the case when $Q^0$ is in the $(2, 2^\prime)$ representation. We then get that  
\beqa
[Q] & = & [(\cos \frac{\theta}{2} + 2 i  J \sin  \frac{\theta}{2}) Q^0] =  \cos \frac{\theta}{2} [Q^0]  \cr
[Q^\prime] & = & [(\cos \frac{\theta}{2} + 2 i  J^\prime \sin  \frac{\theta}{2}) Q^0] =  \cos \frac{\theta}{2} [Q^0] .  
\eeqa
In this case, $[Q]$ and $[Q^\prime]$ can thus both be represented by generators in the $(2, 2^\prime)$ representation, and are in fact equal to each other:
\beq
[Q^\prime] = [Q] .
\eeq

Consider then the case when $Q^0$ is in the $(2^\prime, 2)$ representation. We now get that 
\beqa
[Q] & = & [(\cos \frac{\theta}{2} + 2 i  J \sin  \frac{\theta}{2}) Q^0] =  2 i \sin \frac{\theta}{2} [J Q^0]  \cr
[Q^\prime] & = & [(\cos \frac{\theta}{2} + 2 i  J^\prime \sin  \frac{\theta}{2}) Q^0] =  2 i \sin \frac{\theta}{2} [J^\prime Q^0] .  
\eeqa
In this case, $[Q]$ and $[Q^\prime]$ can thus both be represented by generators in the $(2, 2)$ representation, but they are not equal to each other. Indeed, since
\beq
J^\prime = (\cos \rho - 2 i \hat{J} \sin \rho ) J ,
\eeq
we find that
\beq
[Q^\prime] =  (\cos \rho - 2 i \hat{J} \sin \rho ) [Q] .
\eeq
So $[Q^\prime] = e^{- i \rho} [Q]$ for $Q$ of type $2_{+ \frac{1}{2}}$, whereas  $[Q^\prime] = e^{i \rho} [Q]$ for $[Q]$ of type $2_{- \frac{1}{2}}$.

We can arrange $Q$ or $Q^\prime$ according to their $SO (4)_R$ representation and eigenvalue of $\hat{J}$: 
\beq
\begin{array}{cc}
2_{+ \frac{1}{2}} \nwarrow &  \nearrow 2^\prime_{+ \frac{1}{2}} \cr
2^\prime_{- \frac{1}{2}} \swarrow & \searrow 2_{- \frac{1}{2}} 
\end{array} .
\eeq
(This looks precisely like the arrangement previously given for $Q^0$, but it is important to notice that the labels now refer to $Q$ or $Q^\prime$.) Our results then amount to $Q^\prime$ being given by $Q$ times the following numerical factor:
\beq
\begin{array}{cc}
e^{-i \rho} \nwarrow &  \nearrow 1 \cr
1 \swarrow & \searrow e^{i \rho}
\end{array}
\eeq

We are now ready to formulate the relationships between various amplitudes that follow from supersymmetry. The matrix $\mathcal{M}$ must take the form
\beq
\eqnlab{f}
 \mathcal{M}_{ss'} (|{\bf p}|^2 / T, \theta, \rho) = \de_{ss'} f (|{\bf p}|^2 / T, \theta, \rho) g (\rho, s) ,
\eeq
where $f$ is some function, and the factor $g (\rho, s)$ is given by
\beq
\eqnlab{scattering diamond}
\begin{array}{ccccc}
& & e^{-i \rho} & & \cr
& e^{-i \rho} & & 1 &  \cr
e^{-i \rho} & & 1 & & e^{i \rho} \cr
& 1 & & e^{i \rho} & \cr
& & e^{i \rho} & &  
\end{array}
\eeq
relative to the basis (\ref{sbasis}). All the amplitudes are thus determined \eg by the amplitude $f (|{\bf p}|^2 / T, \theta, \rho)$ for scattering the particles of type $\left| {\bf p}, 4_0 \right>$ into particles of type $\left| {\bf p}^\prime, 4_0 \right>$. This is as far as we get by only using symmetry arguments. To determine the function $f$ requires more information, such as the specific interaction discussed in the next section.

Finally, we remark that, with respect to a basis in which $\frac{1}{\sqrt{2}} (\left| {\bf p}, 1_0 \right> \mp  \left| {\bf p}, 1^\prime_0 \right>)$ have been replaced by the orthonormal vectors $\left| {\bf p}, 1_0 \right>$ and $\left| {\bf p}, 1_0^\prime \right>$, the scattering amplitude is no longer diagonal, but 
\beq
\eqnlab{scattering matrix}
\left(
\begin{array}{cc}
 {}_{\rm in} \left< {\bf p}, 1_0 \right| \left. {\bf p^\prime}, 1_0 \right>_{\rm out}  & {}_{\rm in} \left< {\bf p}, 1_0 \right| \left. {\bf p^\prime}, 1_0^\prime \right>_{\rm out}  \cr
 {}_{\rm in} \left< {\bf p}, 1_0^\prime \right| \left. {\bf p^\prime}, 1_0 \right>_{\rm out}  & {}_{\rm in} \left< {\bf p}, 1_0^\prime \right| \left. {\bf p^\prime}, 1_0^\prime \right>_{\rm out} \cr
 \end{array}
 \right) \sim \left(
 \begin{array}{cc}
 \cos \rho & i \sin \rho \cr
 i \sin \rho & \cos \rho
 \end{array}
 \right) .
 \eeq

Allow us to briefly review the manner in which we will proceed in order to determine the unknown function $f(|{\bf p}|^2 / T, \theta, \rho)$ to lowest order in $|{\bf p}|^2 / T$. Our approach consists of four steps: The first step is to find a suitable model, or interaction, describing the degrees of freedom, being the fields of the tensor multiplet together with the so called bosonic and fermionic string waves $(X,\Th)$. This interaction was constructed in \cite{AFH:2003sc} and is a sum of a Nambu-Goto type term, which involves an integral over the string world-sheet $\Si$ and a Wess-Zumino type term. The latter contains an integral over the world-volume of a Dirac membrane, having $\Si$ as its boundary. We give a short presentation of this action in section \secref{sub action}.

The second step is to rewrite the Wess-Zumino term as an integral over $\Si$ and to collect only the terms contributing to lowest non-trivial order in the scattering processes of our concern. In order to do so we will make various gauge choices and rewrite the degrees of freedom in a suitable manner. These choices are discussed in sections \secref{sub string} and \secref{tm fields}. The end result of these calculations is a rather nice action for the string waves coupled to an on-shell tensor multiplet background, see \Eqnref{taylor action}

The third step is to integrate out the string waves by means of path integrals. This is described in section \secref{effective action}. We then obtain an effective action describing small fluctuations of the tensor multiplet background in the presence of a nearly straight and static infinitely long (\ie BPS saturated cosmic) string.

As the final fourth step, in section \secref{cross section}, we insert the Fourier expansions of the on-shell tensor multiplet fields into the effective action. (The Fourier expansions are thorougly described in section \secref{tm fields}.) It then follows immediately that the sought function $f(|{\bf p}|^2 / T, \theta, \rho)\sim |\pbf|^2/T$, but carries no angular dependence at all, to lowest order in $|{\bf p}|^2 / T$.

We end the paper by obtaining also the differential cross section for scattering of a particle against a string. The result is very simple, see \Eqnref{cs}.

\section{The model}
\seclab{action}
\renewcommand{\thefootnote}{\fnsymbol{footnote}}
\setcounter{footnote}{1}

In this section we review the supersymmetric model of \cite{AFH:2003sc}, describing a self-dual, spinning string coupled to an on-shell (2,0) tensor multiplet background. We start by describing the degrees of freedom living on the string world-sheet. We then move on to the fields of the tensor multiplet and go through their Fourier expansions in some detail. Having done that, it is straight forward to quantize the tensor multiplet and, in particular, obtain the Fock space of one-particle states. Finally, we present the string action and expand it to sufficient order in perturbation theory, when restricting the string to be approximately straight and infinitely long.

Notice that in our model, all interactions take place on the string world-sheet. So, in the absence of a string, the tensor multiplet fields are free. The mere fact that it is possible to construct such interaction terms in a supersymmetric fashion seems to suggest that the tensor multiplet fields do not interact directly with each other. And, as was mentioned in the introduction, this is also what we are to expect of a tensor multiplet, since it is unnatural for a two-form gauge field to couple to particles; a two-form couples naturally to strings. Supersymmetry then implies that all fields should interact only on the string world-sheet.

From now on, we drop the index-free notation of the previous section and make use of $SO(5,1)$ Weyl indices $\alh,\beh=1,...,4$ (see Appendix
\secref{weyl} for details) and $SO(5)$ spinor indices
$\ah, \bh = 1,...,4$. The latter indices can be raised and lowered from the left by the $SO(5)$ invariant tensor $\Om^{\ah\bh} =
-\Om^{\bh\ah}$ and its inverse $\Om_{\ah\bh} = -\Om_{\bh\ah}$. Consistency then requires that $\Om^{\ah\bh} \Om_{\bh \ch} =
\de^{\ah}_{\ch}$. We choose not to use conventional vector indices, although they can sometimes be advantageous, since this is really the way to go when working in the supersymmetric theory.

\subsection{The string}
\seclab{sub string}

The presence of a string in six-dimensional Minkowski spacetime breaks the translational symmetry in the four directions transverse to the string world-sheet,
$\Si$. This gives rise to four Goldstone bosons on $\Si$, which describe the fluctuations of the string in the transverse directions. However, since the string is not necessarily straight, different parts of the string may break different parts of the translational symmetry. This imposes some difficulties in working with the Goldstone bosons, since one would have to consider different sets of Goldstone bosons on different parts of the string. It would be preferable to keep the full Lorentz covariance somehow and describe the whole string using the same set of Goldstone bosons. This can be done by adding two extra bosonic fields on $\Si$, which combine with the Goldstone bosons to form a Lorentz vector $X^{\alh\beh}(\tau,\si)=-X^{\beh\alh}(\tau,\si)$,
where $\tau,\si$ parametrizes $\Si$. The extra degrees of freedom can be fixed by a choice of parametrization of the string world-sheet. Furthermore, the breaking of supersymmetry gives rise to four fermionic degrees of freedom realized by eight Goldstone fermions living on the world-sheet. Analogously to the bosonic case, we add eight extra fermionic fields, which combine with the Goldstone fermions to yield an anti-chiral Lorentz spinor $\Th^{\alh}_{\ah}(\tau,\si)$ being a spinor also under the $SO(5)$ $R$-symmetry group and obeying the following symplectic Majorana condition
\beq
(\Th^{\alh}_{\ah})^* = -C^{*\alh}_{\ph{*}\beh} \Om^{\ah\bh}
\Th^{\beh}_{\bh}.
\eeq
In this equation, $C_{\ph{*}\alh}^{*\beh}$ is the (complex conjugate of the) charge conjugation matrix obeying $C_{\gah}^{\beh} C^{*\gah}_{\ph{*}\alh} = -\de^{\beh}_{\alh}$. We also have that $(\Om^{\ah\bh})^*=-\Om_{\ah\bh}$. The redundancy of fermionic degrees of freedom can be fixed by a local fermionic $\ka$-symmetry of the action \cite{AFH:2003sc}. We will return to this issue shortly. In the remainder of this paper we refer to $X$ and $\Th$ as the bosonic and fermionic string waves, respectively.

In the scattering problem we restrict our attention to an approximately straight and infinitely long string. We then work perturbatively in the parameter $E/\sqrt{T}$, where $E$ is the energy of the incoming tensor multiplet particle and $T$ is the string tension. Let us therefore present a suitable way of describing the string waves for this specific problem. To begin with, we note that a straight infinitely long string pointing in the spatial direction $\nbf$ spontaneously breaks the Lorentz group as
\beq
SO(5,1) \rightarrow SO(1,1) \times SO(4)_{\nbf} \simeq SO(1,1) \times SU(2)
\times SU(2).
\eeq
Here, $SO(4)_{\nbf}$ is the subgroup of spatial transformations that leave the string world-sheet invariant. It is natural to make use of the isomorphism above and introduce the $SU(2)$-indices $\al,\be =1,2$ and $\ald,\bed=1,2$ in such a way that $\alh=(\al,\ald)$. The Lorentz vector $X^{\alh\beh}$ then decomposes into $X^{\al\be} $, $X^{\ald\bed}$ and $X^{\al\bed}$. Without loss of generality we suppose that the string is pointing in the $x^5$-direction, \ie $\nbf=(0,0,0,0,1)$, which means that the string world-sheet $\Si$
fills the entire $x^0x^5$-plane. We then interpret $X^{\al\bed}(\tau,\si)$ as the fluctuations of $\Si$ in the transverse directions $x^1,...,x^4$. Furthermore, we choose the parametrization $\tau=x^0$ and $\si=x^5$. It is convenient to introduce light-cone variables on $\Si$,
\beqa
\si^+ &=& \tau+\si\\
\si^- &=& \tau-\si
\eeqa
and the derivatives
\beqa
\pa_+ &=& \frac{1}{2}\left( \pa_\tau+\pa_\si \right)\\
\pa_- &=& \frac{1}{2}\left( \pa_\tau-\pa_\si \right).
\eeqa
It follows that $\pa_+ \si^+= \pa_- \si^-=1$ and from the
identifications $\tau=X^0$, $\si=X^5$ together with the conventions in
Appendix \secref{l.c. conventions} we find the very useful equalities
\beqa
\pa_+ X^{\al\be} &=& \frac{1}{2}\eps^{\al\be}\\
\pa_- X^{\ald\bed} &=& -\frac{1}{2}\eps^{\ald\bed}\\
\pa_+ X^{\ald\bed} &=& \pa_- X^{\al\be} = 0.
\eeqa
Bearing this is mind, it is natural to separate $X$ in two parts; $X_0$ describing the zero modes and $\Xh$ describing the fluctuations of the string world-sheet. We may then expand the bosonic string waves as
\beq
\eqnlab{x shift}
X^{\alh\beh}(\tau,\si) \equiv X_0^{\alh\beh}(\tau,\si) + 
\frac{1}{\sqrt{T}} \Xh^{\alh\beh}(\tau,\si),
\eeq
where the zero-modes of $X$ are
\beqa
X_0^{\al\bed} &=& 0\\
X_0^{\al\be} &=& \frac{1}{2}\eps^{\al\be}\si^+\\
X_0^{\ald\bed} &=&-\frac{1}{2}\eps^{\ald\bed}\si^-,
\eeqa
because of the parametrization. $\Xh$ describes the fluctuations of the string world-sheet in the transverse directions, hence $\Xh^{\al\be}=\Xh^{\ald\bed}=0$ at all times. The factor $1/\sqrt{T}$ is natural to include, since in the action the kinetic term for $\Xh$ will be multiplied by $T$. We therefore call $\Xh^{\al\bed}$ the canonically normalized bosonic string waves. Eventually, we will see that the string waves are suppressed by precisely a factor $1/\sqrt{T}$ in the scattering processes that we are interested in.

To deal with the world-sheet fermions we first notice that the introduction of a string breaks the $R$-symmetry as
\beq
SO(5) \rightarrow SO(4) \simeq SU(2) \times SU(2).
\eeq
This is because the string tension $T$ is given by the norm of the scalar moduli fields, which are given by a certain $SO(5)$ vector, as discussed in section \secref{kinematics}. (See also \Eqnref{string tension}.) The $SO(4)$ is the subgroup of $SO(5)$ which leaves this vector invariant. Using the isomorphism above, we may decompose the $SO(5)$ spinor index $\ah=1,...,4$ as $\ah=(a,\ad)$, where $a=1,2$ and
$\ad=1,2$ are two different sets of $SU(2)$-indices. It is possible to break the $SO(5)$ $R$-symmetry in such a way that the $SO(5)$ invariant tensor decomposes as $\Om^{a\ad}=0$, $\Om^{ab} \equiv
i\eps^{ab}$ and $\Om^{\ad\bd} \equiv i\eps^{\ad\bd}$, where
\beqa
\eqnlab{eps^ab}
\eps^{ab} \eps_{cb} &=& \de^a_c\\
\eqnlab{eps^adbd}
\eps^{\ad\bd} \eps_{\cd\bd} &=& \de^{\ad}_{\cd}
\eeqa
and $\eps^{12}=1$ for both $\eps^{ab}$ and $\eps^{\ad\bd}$.

The world-sheet fermions $\Th^{\alh}_{\ah}$ then
decompose into $\Th^{\al}_{a}$, $\Th^{\ald}_{\ad}$, $\Th^{\al}_{\ad}$
and $\Th^{\ald}_{a}$. However, we fix the $\ka$-symmetry by choosing
$\Th^{\al}_{\ad} = \Th^{\ald}_{a} = 0$.

Similarly to the bosonic case, we introduce the canonically normalized fermionic string waves as
\beq
\eqnlab{th_hat}
\Th^{\alh}_{\ah}(\tau,\si) \equiv \frac{1}{\sqrt{T}}
\Thh^{\alh}_{\ah}(\tau,\si).
\eeq

\subsection{The tensor multiplet in Fourier space}
\seclab{tm fields}
We start by presenting the tensor multiplet fields in terms of $SO(5,1)$ Weyl indices and $SO(5)$ spinor indices. The scalar fields, transforming as an $SO(5)$ vector, are then $\phi^{\ah \bh} = -\phi^{\bh
  \ah}$, subject to the algebraic constraint $\phi^{\ah\bh} \Om_{\ah\bh} = 0$. We know that the string tension is given by the norm of the vacuum expectation values of the scalar fields:
\beq
\eqnlab{string tension}
T=| \big< \phi^{\ah\bh} \big> |.
\eeq
This can be calculated
as $\big< \phi^{\ah\bh} \big> = \lim_{x\rightarrow \infty}\phi^{\ah\bh}(x)$, which we choose to be independent of the direction in which the limit is taken. We can then introduce a constant $SO(5)$ unit vector $\phi_\infty^{\ah\bh} \equiv \big< \phi^{\ah\bh} \big> /T$, such that the inner product is $\Om_{\ah\ch}\Om_{\bh\dha} \phi_\infty^{\ah\bh} \phi_\infty^{\ch\dha}/4 = 1$. It is useful to rewrite the scalar fields as
\beq
\phi^{\ah\bh}_{\sss{\mathrm{TOT}}}(x) = T \phi_\infty^{\ah\bh} + \phi^{\ah\bh}(x),
\eeq
where $\phi^{\ah\bh}(x)$ now denotes the dynamic part (with zero vacuum expectation value). The four freely adjustable parameters in $\phi_\infty^{\ah\bh}$ together with $T$ constitute the moduli of the $A_1$ version of (2,0) theory. We may choose $\phi_\infty^{\ah\bh}$ such that
\beqa
\phi_\infty^{ab} &=& i \eps^{ab}\\
\phi_\infty^{\ad\bd} &=& -i \eps^{\ad\bd}\\
\phi_\infty^{a\bd} &=& 0
\eeqa
under the decomposition of $\ah$ into $a$ and $\ad$, discussed at the end of the previous subsection. One may explicitly check that $\phi_\infty^{\ah\bh}\Om_{\ah\bh}=0$, as required. The two-form gauge field is $b^{\alh}_{\beh}$, where $b^{\alh}_{\alh}=0$. Its three-form field strength $h$ separates into a self-dual part $h_{\alh\beh}=h_{\beh\alh}$ and an anti self-dual part $h^{\alh\beh}=h^{\beh\alh}$,
\beqa
\eqnlab{h}
h_{\alh\beh} &=& \pa_{\alh\gah}b^{\gah}_{\beh} +
\pa_{\beh\gah}b^{\gah}_{\alh} \\
h^{\alh\beh} &=& \pa^{\alh\gah}b^{\beh}_{\gah} +
\pa^{\beh\gah}b^{\alh}_{\gah}.
\eeqa
As already mentioned, only $h_{\alh\beh}$ is part of the
tensor multiplet. Hence, we have that $h^{\alh\beh}=0$. The field strength obeys the Bianchi identity
\beq
\eqnlab{bianchi}
\pa^{\alh\gah} h_{\alh\beh} - \pa_{\alh\beh} h^{\alh\gah} = 0.
\eeq
Finally, the fermionic fields are $\psi^{\alh}_{\ah}$, subject to the
symplectic Majorana condition
\beq
\eqnlab{majorana}
(\psi_{\alh}^{\ah})^* = C_{\alh}^{\beh} \Om_{\ah\bh} \psi_{\beh}^{\bh}.
\eeq

The dynamics of a single tensor multiplet background is governed by a free action. However, it is well known that there does not exist a Lagrangian description for a self-dual three-form in six dimensions \cite{Witten:1997}. We can remedy this by including also the anti self-dual part of $h$ as a 'spectator field'. (An alternative solution to this problem can be found in \cite{Pasti:1996}.) Up to an overall constant, the supersymmetric and $SO(5)$ invariant action for a free tensor multiplet in six dimensions can then be written as \cite{AFH:2003sc}
\beq
\eqnlab{tm action}
S_{\sss TM} = \int d^6 x \left[ - \pa_{\alh \beh} \phi^{\ah\bh}
  \pa^{\alh \beh} \phi_{\ah\bh} + 2
h_{\alh \beh} h^{\alh \beh} - 4i \Om_{\ah\bh} \psi^{\ah}_{\alh}
\pa^{\alh \beh} \psi^{\bh}_{\beh} \right],
\eeq
and the supersymmetry transformations are
\beqa
\eqnlab{tm susy phi}
\de \phi^{\ah\bh} &=& i \left( \Om^{\ah\ch} \eta^{\alh}_{\ch} \psi^{\bh}_{\alh} - \Om^{\bh\ch} \eta^{\alh}_{\ch} \psi^{\ah}_{\alh} - \frac{1}{2} \Om^{\ah\bh} \eta^{\alh}_{\ch} \psi^{\ch}_{\alh}  \right)\\
\de \psi^{\ah}_{\alh} &=& \Om^{\ah\bh} \eta^{\beh}_{\bh} h_{\alh\beh} + 2 \eta^{\beh}_{\bh} \pa_{\alh\beh} \phi^{\ah\bh}\\
\eqnlab{tm susy h}
\de h_{\alh\beh} &=& -i \eta^{\gah}_{\ah} \left( \pa_{\alh\gah} \psi^{\ah}_{\beh} + \pa_{\beh\gah} \psi^{\ah}_{\alh} \right),
\eeqa
where $\eta$ is a constant fermionic parameter.

We will now derive the equations of motion that follow from this action and write down the general solutions to these by means of Fourier expansions. The seemingly arbitrary constants in the following expansions are chosen such that upon a canonical quantization, the Hamiltonian that follows from the free action \eqnref{tm action} is correct.

\subsubsection{The scalar fields}
We start by varying the action with respect to the scalar fields. This yields the Klein-Gordon equation of motion:
\beq
\pa_{\alh \beh} \pa^{\alh \beh} \phi^{\ah\bh} = 0,
\eeq
which has the general solution
\beq
\eqnlab{phi}
\phi^{\ah\bh}(x) = \frac{1}{4}\int \frac{d^5p}{(2\pi)^{5/2}}
\frac{1}{\sqrt{|\pbf|}}
\left(a^{\ah\bh}(\pbf)e^{ip\cdot x} + a^{*
  \ah\bh}(\pbf)e^{-ip\cdot x} \right).
\eeq
The momentum $p^{\alh\beh}$ is light-like (\ie $p^{\alh\beh}
p_{\alh\beh} = 0$) since the scalar fields are massless. The
coefficients $a^{\ah\bh}(\pbf)$ and $a^{* \ah\bh}(\pbf) \equiv
\Om^{\ah\ch} \Om^{\bh\dha} a^*_{\ch\dha}(\pbf)$ are functions in
Fourier space that parametrize the solutions. They are subject to the
same tracelessness condition as $\phi^{\ah\bh}$, \ie
$\Om_{\ah\bh} a^{\ah\bh} = \Om^{\ah\bh} a^*_{\ah\bh} = 0.$ It is
easy to check that $\phi^{\ah\bh}$ obeys the reality condition
\beq
(\phi^{\ah\bh})^* = \Om_{\ah\ch} \Om_{\bh\dha} \phi^{\ch\dha} \equiv
\phi_{\ah\bh}.
\eeq
It will be useful to define also a real scalar field $\phi_\|$ as being the projection of $\phi^{\ah\bh}$ on the specific $SO(5)$ unit vector $\phi_\infty^{\ah\bh}$, \ie
\beq
\phi_\| = \frac{1}{4} \phi_\infty^{\ah\bh} \phi_{\ah\bh}.
\eeq
Its Fourier expansion becomes
\beq
\phi_\|(x) = \frac{1}{4}\int \frac{d^5p}{(2\pi)^{5/2}}
\frac{1}{\sqrt{|\pbf|}}
\left(a_{\phi_\|}(\pbf)e^{ip\cdot x} + a^*_{\phi_\|}(\pbf)e^{-ip\cdot x} \right),
\eeq
where $a_{\phi_\|}(\pbf) = \frac{1}{4} \phi_\infty^{\ah\bh} a_{\ah\bh}(\pbf)$. This particular scalar field describes the polarization $1_0$ of section \secref{kinematics}. The other four polarizations $4_0$ are contained in $\phi^{a\bd}$.

\subsubsection{The fermionic fields}
The same analysis for $\psi$ leads to the Dirac equation of motion
\beq
\eqnlab{psi eq}
\pa^{\alh\beh} \psi^{\bh}_{\beh} = 0.
\eeq
It is straight forward to show that the solutions to the Dirac
equation are chiral spinors under the $SO(4)$ little
group of transformations that leave invariant the momentum of the
spinor. Such spinors carry two polarizations, which were labeled $+\frac{1}{2},-\frac{1}{2}$ in the previous section. Since each spinor also has an $SO(5)_R$ spinor index taking four values, we end up with eight different fermionic polarizations. It now follows that the general form of $\psi$ is
\beq
\psi^{\ah}_{\alh}(x) = \frac{1}{2} \int \frac{d^5 p}{(2\pi)^{5/2}}
\sum_s \left( u_{\alh}(\pbf,s)a^{\ah}(\pbf,s) e^{ip\cdot x} +
v_{\alh}(\pbf,s)a^{* \ah}(\pbf,s) e^{-ip\cdot x}
\right),
\eeq
where the summation is over $s=+\frac{1}{2},-\frac{1}{2}$. Comparing with section \secref{kinematics}, we have that $a^a(\pbf,s)$ and $a^{\ad}(\pbf,s)$ correspond to the polarizations $2'_s$ and $2_s$, respectively. We also have that $a^{* \ah} \equiv \Om^{\ah\bh}a^*_{\bh}$ and that the
momentum $p^{\alh\beh}$ is light-like. Furthermore, the fields $u_{\alh}(\pbf,s)$ and $v_{\alh}(\pbf,s)$ independently span the two-dimensional vector space of linearly independent solutions to the Dirac equation, hence
\beqa
p^{\alh\beh} u_{\beh}(\pbf,s) &=& 0\\
p^{\alh\beh} v_{\beh}(\pbf,s) &=& 0.
\eeqa
However, they are related to each other by the symplectic Majorana condition \eqnref{majorana}, which implies that
\beqa
u^*_{\alh}(\pbf,s) &=& C_{\alh}^{\beh} v_{\beh}(\pbf,s) \\
v^*_{\alh}(\pbf,s) &=& -C_{\alh}^{\beh} u_{\beh}(\pbf,s).
\eeqa
The minus sign in one of these relations is needed in order for
$(u_{\alh})^{* *} = u_{\alh}$, since the charge
conjugation matrix obeys $C^{\beh}_{\gah} C^{*\gah}_{\ph{*}\alh} = -
\de^{\beh}_{\alh}$. We also have that
\beqa
(a^{\ah}(\pbf,s))^* &=& a^*_{\ah}(\pbf,s)\\
(a^*_{\ah}(\pbf,s))^* &=& a^{\ah}(\pbf,s).
\eeqa
Because of the reality conditions on $u$ and $v$, we can write the inner product on the space of solutions to the Dirac equation as
\beq
\frac{\pbf^{\alh\beh}}{|\pbf|}u_{\alh}(\pbf,s) v_{\beh}(\pbf,s') = - \frac{1}{2} \de_{ss'}.
\eeq
The functions $u_{\alh}(\pbf,s)$ and $v_{\alh}(\pbf,s)$ for a generic momentum $\pbf$ are obtained from $u_{\alh}(\nbf,s)$ and $v_{\alh}(\nbf,s)$ for the specific momentum $\nbf = (0,0,0,0,1)$ as
\beqa
u_{\alh}(\pbf,s) &=& L_{\alh}^{\ph{\alh}\beh}(\pbf) u_{\beh}(\nbf,s)\\
v_{\alh}(\pbf,s) &=& L_{\alh}^{\ph{\alh}\beh}(\pbf) v_{\beh}(\nbf,s).
\eeqa
where $L_{\alh}^{\ph{\alh}\beh}(\pbf)$ is the transformation induced by the standard rotation $L(\pbf)$ when acting on a spinor. This was described in the previous section, see \Eqnref{spinor transf}.

Chirality implies that the reference basis functions only have non-zero components for the dotted indices. Furthermore, they obey
\beqa
(J^{12})_{\alh}^{\ph{\alh}\beh} u_{\beh}(\nbf,+\frac{1}{2}) &=& +\frac{1}{2} u_{\beh}(\nbf,+\frac{1}{2})\\
(J^{12})_{\alh}^{\ph{\alh}\beh} u_{\beh}(\nbf,-\frac{1}{2}) &=& -\frac{1}{2} u_{\beh}(\nbf,-\frac{1}{2})
\eeqa
where $(J^{12})_{\alh}^{\ph{\alh}\beh} = -\frac{i}{2}\Ga^1_{\gah\alh} \Ga^{2 \gah\beh}$ is the generator of spatial rotations in the $x^1 x^2$-plane.

\subsubsection{The chiral gauge field}
Finally, we turn to $h$: Varying the action with respect to $b$ leads to the following equation of motion:
\beq
\pa^{\alh\gah} h_{\alh\beh} + \pa_{\alh\beh} h^{\alh\gah} = 0.
\eeq
Applying the self-duality constraint, the equation of motion
coincides with the Bianchi identity \eqnref{bianchi}
\beq
\pa^{\alh\beh} h_{\beh\gah} = 0,
\eeq
which is the Maxwell equation for a self-dual three-form. We want to write down the most general chiral two-form $b$ whose field strength obeys this equation. We note that a chiral two-form in six dimensions carries three degrees of freedom, or polarizations. (These were labeled $1'_0,1'_{+1},1'_{-1}$ in section \secref{kinematics}.) As it stands, $b_{\alh}^{\beh}$ obeying $b_{\alh}^{\alh} = 0$ has fifteen components. We thus need to gauge fix the two-form: We start by choosing the Lorentz gauge
\beq
\eqnlab{lorentz}
\pa_{\gah \alh} b^{\gah}_{\beh} - \pa_{\gah \beh} b^{\gah}_{\alh}
= 0.
\eeq
This fixes five of the components of $b$. (In vector index notation,
this gauge choice is written $\pa^\mu b_{\mu\nu}=0$.) We then also choose the condition that
\beq
\eqnlab{gauge 2}
(\Ga^0)_{\gah\alh} b^{\gah}_{\beh} - (\Ga^0)_{\gah\beh}
b^{\gah}_{\alh}=0.
\eeq
(This amounts to setting $b_{0\mu}=0$.) Hence, this seems to fix
another five components of $b$. But in fact it only fixes four, because one of the five
conditions overlaps with one of the Lorentz gauge conditions. We are
then left with six components of $b$, but these are halved to three by
restricting $b$ to be chiral, \ie
\beq
h^{\alh\beh} = \pa^{\alh\gah} b^{\beh}_{\gah} +
\pa^{\beh\gah} b^{\alh}_{\gah} = 0.
\eeq
The three remaining degrees of freedom in $b$ are to be interpreted
as the three different polarizations of the gauge field. It is thus natural to expand $b$ as
\beq
b_{\alh}^{\beh}(x) = \frac{1}{2} \int \frac{d^5p}{(2\pi)^{5/2}}
\frac{1}{\sqrt{|\pbf|}} \sum_s \left(
a(\pbf,s)u_{\alh}^{\beh}(\pbf,s) e^{ip\cdot x} +
a^*(\pbf,s)v_{\alh}^{\beh}(\pbf,s) e^{-ip\cdot x} \right),
\eeq
where the summation is over $0,+1,-1$, and $p^{\alh\beh}$ is light-like. (Hence, $a(\pbf,s)$ correspond to the polarizations $1'_s$.) The Fourier fields $u_{\alh}^{\beh}(\pbf,s)$ and $v_{\alh}^{\beh}(\pbf,s)$
independently span the three-dimensional vector space of solutions to
the equation of motion in Fourier space that are consistent with
the gauge conditions and the chirality property. They are related to each other by a reality condition which makes it possible to write the inner product in this vector space as
\beq
u_{\alh}^{\beh}(\pbf,s)v^{\alh}_{\beh}(\pbf,s') = \frac{1}{2} \de_{ss'}.
\eeq
The fields $u^{\alh}_{\beh}(\pbf,s)$ and $v^{\alh}_{\beh}(\pbf,s)$ for a generic momentum $\pbf$ are obtained from $u^{\alh}_{\beh}(\nbf,s)$ and $v^{\alh}_{\beh}(\nbf,s)$ for the specific momentum $\nbf$ as
\beqa
u_{\alh}^{\beh}(\pbf,s) &=& L_{\alh}^{\ph{\alh}\gah}(\pbf) L^{\beh}_{\ph{\beh}\deh}(\pbf) u_{\gah}^{\deh}(\nbf,s)\\
v_{\alh}^{\beh}(\pbf,s) &=& L_{\alh}^{\ph{\alh}\gah}(\pbf) L^{\beh}_{\ph{\beh}\deh}(\pbf) v_{\gah}^{\deh}(\nbf,s).
\eeqa
As in the case of the fermions, chirality (or, likewise, the self-duality of $h$) implies that the reference basis functions are non-zero only for dotted indices, \ie $u_{\al}^{\be}(\nbf,s) = u_{\al}^{\bed}(\nbf,s) = u_{\ald}^{\be}(\nbf,s) = 0$. (For an anti-chiral gauge field, the functions $u(\nbf,s)$ would have had only undotted indices.) Most importantly, we have that
\beq
(J^{12})_{\ald}^{\ph{\alh}\gah} u_{\gah}^{\beh}(\nbf,s) +  (J^{12})_{\ph{\beh}\gah}^{\beh} u_{\alh}^{\gah}(\nbf,s) = s u_{\alh}^{\beh}(\nbf,s).
\eeq
In the corresponding relations for $v_{\alh}^{\beh}(\nbf,s)$, the right hand side is multiplied by minus one. In one of the calculations to be done, one must also use that
\beq
(J^{12})_{\ald}^{\ph{\ald}\gad} u_{\gad}^{\bed}(\nbf,0) = -\frac{1}{4} \de_{\ald}^{\bed}.
\eeq

\subsubsection{Quantizing the tensor multiplet}
\seclab{quantization}

To make further connection to section \secref{kinematics}, we show how to quantize a free tensor multiplet. The various functions $a(\pbf)$ and their complex conjugates $a^*(\pbf)$ then become annihilation and creation operators ($a(\pbf)$ and $a^\dagger(\pbf)$) acting on a Fock space of particle states. We impose the following commutation relations on these ladder operators:
\beqa
\left[ a_{\phi_\|}(\pbf), a^\dagger_{\phi_\|}(\pbf') \right] &=& \de^{(5)}(\pbf-\pbf')\\
\left[ a^{a\bd}(\pbf), a^\dagger_{c\dd}(\pbf') \right] &=& 2\de^{(5)}(\pbf-\pbf')\de^a_c \de^{\bd}_{\dd}\\
\left[ a^{\ah}(\pbf,s), a^\dagger_{\bh}(\pbf',s') \right] &=& \de^{(5)}(\pbf-\pbf')\de^{\ah}_{\bh} \de_{ss'}, \quad s={\sss \pm\frac{1}{2}}\\
\left[ a(\pbf,s), a^\dagger(\pbf',s') \right] &=& \de^{(5)}(\pbf-\pbf') \de_{ss'}, \quad s= \pm 1, 0.
\eeqa
The Fock space of particle states is then constructed by acting on the vacuum $\left| 0 \right>$ with the creation operators. In particular, comparing with the one-particle states of (\ref{sbasis}) we have the following relations
\beqa
\left| \pbf, 1_0 \right> &\leftrightarrow&  a^\dagger_{\phi_\|}(\pbf) \left| 0 \right> \\
\left| \pbf, 4_0\right> &\leftrightarrow&  \frac{1}{\sqrt{2}} a^\dagger_{a\bd}(\pbf) \left| 0 \right>\\
\left| \pbf, 1'_s\right> &\leftrightarrow&  a^\dagger(\pbf,s) \left| 0 \right>, \quad s= \pm 1, 0\\
\left| \pbf, 2'_s\right> &\leftrightarrow&  a^\dagger_a(\pbf,s) \left| 0 \right>, \quad s= {\sss \pm\frac{1}{2}}\\
\left| \pbf, 2_s\right> &\leftrightarrow& a^\dagger_{\ad}(\pbf,s) \left| 0 \right>, \quad s= {\sss \pm\frac{1}{2}} .
\eeqa

\subsection{The action}
\seclab{sub action}

In this subsection, we present shortly the action of \cite{AFH:2003sc}. We continue with a description of how it can be rewritten by means of a perturbative expansion. Finally, we give the end result of this expansion. At first, the action might seem a bit complicated, but it will simplify greatly when put in the expanded form, tailor made for our scattering problem.

Let us begin by introducing a superspace with both bosonic coordinates $x^{\alh\beh} = - x^{\beh\alh}$ and fermionic coordinates $\theta^{\alh}_{\ah}$ \cite{Howe:1983}. The bosons and fermions on the string world-sheet $(X,\Th)$ may then be interpreted as the coordinates of the string in this superspace. The action for a string coupled to an on-shell tensor multiplet background is
\beq
\eqnlab{action}
S = -\int_\Si d^2\si \sqrt{\Om_{\ah\ch}\Om_{\bh\dha} \Phi^{\ah\bh}(X,\Th)\Phi^{\ch\dha}(X,\Th)} \sqrt{-G} + \int_D {}^*F.
\eeq
Here, $\Phi^{\ah\bh}(X,\Th)$ is a certain superfield evaluated at $\Si$. Explicitly, we have
\beqa
\Phi^{\ah\bh}(x,\theta) &=& \phi^{\ah\bh}_{\sss{\mathrm{TOT}}} - i\theta^{\alh}_{\ch}\Big( \Om^{\ch\ah} \psi^{\bh}_{\alh} + \Om^{\bh\ch} \psi^{\ah}_{\alh} + \frac{1}{2} \Om^{\ah\bh} \psi^{\ch}_{\alh} \Big) +\nn\\ &&{}+ i\theta^{\alh}_{\ch} \theta^{\beh}_{\dha} \Big( h_{\alh\beh} \Big( \Om^{\dha\ah}\Om^{\bh\ch} + \frac{1}{4} \Om^{\dha\ch}\Om^{\ah\bh}  \Big) -\Om^{\dha\ah} \pa_{\alh\beh} \phi^{\bh\ch} - \Om^{\dha\bh} \pa_{\alh\beh} \phi^{\ch\ah} \Big) +\nn\\
&&{}+ \ordo(\theta^3).
\eeqa
A priori, the final term of this superfield is of order $\theta^{16}$.
However, when evaluating it on $\Si$ each $\Th$ is suppressed by a factor $T^{-1/2}$ (see \Eqnref{th_hat}), so we need not bother with terms of higher order in our perturbative approach.

Furthermore, $G$ is the determinant of the induced metric on $\Si$,
\beq
G_{ij} = \frac{1}{2} \eps_{\alh\beh\gah\deh} \left( \pa_i X^{\alh\beh}
+ i\Om^{\ah\bh} \Th^{[\alh}_{\ah} \pa_i\Th^{\beh]}_{\bh} \right)
\left( \pa_j X^{\gah\deh} + i\Om^{\ch\dha} \Th^{[\gah}_{\ch}
  \pa_i\Th^{\deh]}_{\dha} \right),
\eeq
where $i,j=+,-$.

In the last term of the action, the integral is over the world-volume of a 'Dirac
membrane' $D$, which has $\Si$ as its boundary, \ie $\pa D= \Si$ \cite{Deser:1998}. This is analogous to the Dirac string in four dimensions. The integrand is the pull-back to $D$ of a certain closed super-three-form (first introduced in \cite{Grojean:1998}):
\beqa
F &=& \frac{1}{3} e^{\gah_1\gah_2}\we e^{\beh_1\beh_2}\we e^{\alh_1\alh_2} \eps_{\alh_2\beh_2\gah_1\gah_2} H_{\alh_1\beh_1}+\nn\\
&&{}+ i e^{\gah_1\gah_2}\we e^{\beh_1\beh_2}\we d \theta^{\alh}_{\ah} \eps_{\beh_2\alh\gah_1\gah_2} \Psi^{\ah}_{\alh}+\nn\\
&&{}+ i e^{\gah_1\gah_2}\we d \theta^{\beh}_{\ah}\we d \theta^{\alh}_{\ah} \eps_{\alh\beh\gah_1\gah_2} \Phi^{\ah\bh}
\eeqa
where
\beq
e^{\alh\beh} = d x^{\alh\beh}
+ i\Om^{\ah\bh} \theta^{[\alh}_{\ah} d\theta^{\beh]}_{\bh},
\eeq
and $dx^{\alh\beh}$, $d\theta^{\alh}_{\ah}$ are superspace
differentials. Furthermore, $\Psi^{\ah}_{\alh}$ is a superfield obtained from $\Phi^{\ah\bh}$ as
\beq
\Psi^{\ah}_{\alh} = -\frac{2i}{5} \Om_{\ah\bh} \left( \pa^{\ah}_{\alh} + i \Om^{\ah\ch} \theta^{\gah}_{\ch} \pa_{\alh\gah} \right) \Phi^{\bh\ch}
\eeq
with $\psi^{\ah}_{\alh}$ as its lowest component. Finally, $H_{\alh\beh}$ is a superfield obtained from $\Psi^{\ah}_{\alh}$ as
\beq
H_{\alh\beh} = \frac{1}{4} \Om_{\ah\bh} \left( \pa^{\ah}_{\alh} + i \Om^{\ah\ch} \theta^{\gah}_{\ch} \pa_{\alh\gah} \right) \Psi^{\bh}_{\beh}
\eeq
with $h_{\alh\beh}$ as its lowest component. The fact that the super-three-form $F$ is closed implies that it should be possible to rewrite the final term in the action as an integral over $\Si$ instead of over $D$ by means of Stokes' theorem. This is indeed what we will do, but we will not be able to preserve the full Lorentz covariance.

Using the free equations of motion for the tensor multiplet fields, it
follows that the two terms in the action \eqnref{action} are both supersymmetric, with
\beqa
\eqnlab{x susy}
\de X^{\alh\beh} &=& i \Om^{\ah\bh} \eta^{[\alh}_{\ah} \Th^{\beh]}_{\bh}\\
\eqnlab{theta susy}
\de \Th^{\alh}_{\ah} &=& - \eta^{\alh}_{\ah}
\eeqa
and the fields of the tensor multiplet transforming according to \Eqsref{tm susy phi}-\eqnref{tm susy h}. We recall that $\eta$ is a constant fermionic parameter.  (Notice that \Eqsref{x susy} and \eqnref{theta susy} are not the transformation properties for the canonically normalized string waves.) The sum of the two terms is also invariant under the following $\ka$-transformations
\beqa
\eqnlab{x kappa}
\de X^{\alh\beh} &=& i \Om^{\ah\bh} \ka^{[\alh}_{\ah} \Th^{\beh]}_{\bh}\\
\eqnlab{theta kappa}
\de \Th^{\alh}_{\ah} &=& \ka^{\alh}_{\ah},
\eeqa
where $\ka$ is a local fermionic parameter. It is subject to the following constraint
\beq
\eqnlab{kappa constraint}
\Ga^{\alh}_{\ph{\alh}\beh} \ka^{\beh}_{\ah} = \ga_{\ah}^{\ph{\ah}\bh} \ka^{\alh}_{\bh},
\eeq
where
\beqa
\Ga^{\alh}_{\ph{\alh}\beh} &=& \frac{1}{\sqrt{-G}} \eps^{ij} \Big( \pa_i X^{\alh\gah} + i \Om^{\ah\ch} \Th^{[\alh}_{\ah} \pa_i \Th^{\gah]}_{\ch} \Big) \Big( \pa_j X^{\deh\epsh} + i \Om^{\dha\eh} \Th^{[\deh}_{\dha} \pa_j \Th^{\epsh]}_{\eh} \Big) \eps_{\beh\gah\deh\epsh}\\
\ga_{\ah}^{\ph{\ah}\bh} &=& \frac{1}{\sqrt{\frac{1}{4}\Phi^{\ch\dha} \Phi_{\ch\dha}}} \Om_{\ah\eh} \Phi^{\eh\bh}.
\eeqa
It is clear that $\Ga^{\alh}_{\ph{\alh}\alh} = \ga_{\ah}^{\ph{\ah}\ah} = 0$. Furthermore, one can show that $\Ga^{\alh}_{\ph{\alh}\beh} \Ga^{\beh}_{\ph{\beh}\gah} = \de^{\alh}_{\gah}$ and $\ga_{\ah}^{\ph{\ah}\bh} \ga_{\bh}^{\ph{\bh}\ch} = \de^{\ch}_{\ah}$. This means that we can use the $\ka$-symmetry to eliminate half of the components of $\Th$, as argued in section \secref{sub string}.

Despite the rather nice form of \Eqnref{action}, the action is very complicated when writing out the superfields explicitly. In order for us to read off any scattering amplitudes, we need to massage it
somehow, and we let perturbation theory guide us.

To begin with, we choose to consider a straight and infinitely long string, pointing in the $x^5$-direction. We then apply the specific parametrization and $\ka$-fixing of section \secref{sub string}. We
also do the change of variables discussed in that section, $X \rightarrow X_0 + \Xh/\sqrt{T}$ and $\Th \rightarrow \Thh /\sqrt{T}$. Having done all that, we Taylor expand the fields of the tensor multiplet. Consider the gauge field as an example:
\beq
b^{\alh}_{\beh}(X) = b^{\alh}_{\beh}(X_0) +
2\frac{1}{\sqrt{T}}\pa_{\ga\ded} b^{\alh}_{\beh}(x)\Big|_{x=X_0}
\Xh^{\ga\ded} + \ordo(T^{-1}).
\eeq

To collect no more than the terms relevant for our scattering problem,
let us pause here for a moment and discuss the dynamics of a typical scattering process: The in-state consists of a tensor multiplet particle with energy $E$, and a straight
string at rest with string tension $T$. When the particle hits
the string it will be absorbed and a number of string waves will be
excited. However, each string wave will be suppressed by a factor
$E/\sqrt{T}$. (Or, rather, the probability amplitude for creating $N$ string waves will be suppressed by a factor $(E/\sqrt{T})^N$.) Hence, the leading contribution to the scattering
amplitude comes from a process where only one string wave is
excited. In this process, both the momentum parallel to the string and
the energy of the incoming particle will be preserved. It follows that
the string wave cannot be on-shell, and that this is not a stable state. Hence,
the string wave will only live for a short while and then emit another
tensor multiplet particle. This latter process will be suppressed by an additional factor
$E/\sqrt{T}$. So the total amplitude for one-particle to one-particle scattering via string wave excitations will be suppressed by at least a factor $E^2/T$. Any such process comes from terms in the action which are linear in tensor multiplet fields and linear in
string waves. They will include a factor $T^{-1/2}$. Note that we will
not have to take pure string wave interactions into account, since
only one string wave at a time will be excited, and therefore it has no other
string wave to interact with.

However, there may also be terms in the action which are bilinear in tensor multiplet fields, but that do not involve any string waves. They correspond to direct scattering of a particle against the string. Since no string waves are excited, these terms can be suppressed by a factor $T^{-1}$ and still contribute to the scattering amplitude to the same order in perturbation theory as the ones just described.

Hence, we are to collect all terms in the action of order $T^{-1/2}$ which are linear in tensor multiplet fields and linear in string waves, but also all terms
of order $T^{-1}$ that are quadratic in tensor multiplet fields, but
lacking string waves. We are then sure to find all terms contributing
to the desired scattering amplitudes to order $E^2/T$. Doing this is quite laborous, but not very difficult. We therefore choose to skip all the details and instead present the final form of these calculations:
\beqa
S &=&  \int_\Si d^2\si \Big[ 4\pa_+\Xh^{\al\bed}\pa_-\Xh_{\al\bed} +
  2\eps_{\al\be}\eps^{ab} \Thh^\al_a\pa_+\Thh^\be_b -
  2\eps_{\ald\bed}\eps^{\ad\bd}
  \Thh^{\ald}_{\ad}\pa_-\Thh^{\bed}_{\bd}+\nn\\
&&{}+ \frac{1}{\sqrt{T}}\Big(
  - 2\pa_{\al\bed}\phi_\|(X_0)\Xh^{\al\bed} + h_{\al\bed}(X_0)
  \Xh^{\al\bed}-  i\psi^a_\al(X_0) \Thh^\al_a +
  i\psi^{\ad}_{\ald}(X_0) \Thh^{\ald}_{\ad} \Big) -\nn\\
\eqnlab{taylor action}&&{}- \frac{1}{4T}
  \phi^{a\bd}(X_0) \phi_{a\bd}(X_0) \Big] +...
\eeqa
The dots in this action indicate terms that contribute to the scattering
amplitudes to higher orders in the parameter $E/\sqrt{T}$. We recall that $\phi_\|$ is defined as the projection of $\phi^{\ah\bh}$ on the $SO(5)_R$ unit vector $\phi_\infty^{\ah\bh}$.

The first line in this action is evidently the kinetic terms for the string waves, then follow the interaction terms. We see that both types of terms discussed above are present. Finally, we stress that we have not yet applied the gauge conditions on $b$. However, we have made use of the self-duality of the field strength $h$.

\section{Scattering}
\seclab{scattering}
\renewcommand{\thefootnote}{\fnsymbol{footnote}}
\setcounter{footnote}{1}

In this section we obtain an effective action describing small fluctuations of the background in the presence of a nearly BPS saturated string. It is obtained by integrating out the string waves from the action in \Eqnref{taylor action}. To lowest order in perturbation theory, we can read off the desired scattering amplitudes from this effective action. By quantizing the tensor multiplet, one may also to calculate the $T$-matrix from it (recall that $S=1+iT$). We end this section by obtaining an expression for the differential cross sections, which turns out to be remarkably simple.

\subsection{The effective action}
\seclab{effective action}

The effective action is defined by
\beq
\eqnlab{s_eff def}
e^{iS_{\sss{eff}}}  = \frac{Z[S]}{Z[S_0]},
\eeq
where
\beq
\eqnlab{Z}
Z[S] = \int \mathcal{D}\Xh^{\al\bed} \mathcal{D}\Thh^\al_a \mathcal{D}\Thh^{\ald}_{\ad} e^{iS}.
\eeq
 In this notation, $S$ is the action \eqnref{taylor action} and $S_0$ is the part of the action that involves only the kinetic terms for the string waves. To be able to integrate out the string waves from the action, we want to complete the square in it. In order to do so, we introduce the string wave propagators
\beqa
D_F(\si,\si') &=& i \int \frac{d^2k}{(2\pi)^2} \frac{e^{ik\cdot (\si-\si')}}{k^2+i\eps}\\
S^+_F(\si,\si') &=& - \int \frac{d^2k}{(2\pi)^2} \frac{e^{ik\cdot (\si-\si')}}{k_+ +i\eps}\\
S^-_F(\si,\si') &=& - \int \frac{d^2k}{(2\pi)^2} \frac{e^{ik\cdot (\si-\si')}}{k_- +i\eps}
\eeqa
where $\eps$ is a small parameter. Notice that we use a notation in which $\si$ denotes the set $(\si^0,\si^1)$ and $\si'=(\si'^0,\si'^1)$ is another set of parameters on the string world-sheet. Furthermore, $k=(k^0,k^1)$ are the momentum space variables for the string waves. It follows that
\beqa
k_+ = \frac{1}{2} (k_0 + k_1)\\
k_- = \frac{1}{2} (k_0 - k_1)
\eeqa
from our conventions in appendix \secref{l.c. conventions}. One can check that
\beqa
\eqnlab{df der}
\pa^2 D_F(\si-\si')&=& -i\de^{(2)}(\si-\si')\\
\pa_+ S^+_F(\si-\si')&=& -i\de^{(2)}(\si-\si')\\
\eqnlab{sf- der}
\pa_- S^-_F(\si-\si')&=& -i\de^{(2)}(\si-\si'),
\eeqa
where the differential operators (acting on $\si$) are the two-dimensional Klein-Gordon operator together with the chiral and anti-chiral Dirac operators respectively. We now introduce the following shifts of variables in the action
\beqa
\Xh'^{\al\bed}(\si) &=& \Xh^{\al\bed}(\si) - \frac{i}{2\sqrt{T}} \eps^{\al\ga} \eps^{\bed\ded} \int d^2\si' D_F(\si-\si') \big( h_{\ga\gad}(\si') - 2\pa_{\ga\gad} \phi_\| (\si') \big) \qquad\\
\Thh'^\al_a(\si) &=& \Thh^\al_a(\si) - \frac{1}{4\sqrt{T}} \eps^{\al\be} \eps_{ab} \int d^2\si' S_F^+(\si-\si') \psi^b_\be(\si') \\
\Thh'^{\ald}_{\ad}(\si) &=& \Thh^{\ald}_{\ad}(\si) - \frac{1}{4\sqrt{T}} \eps^{\ald\bed} \eps_{\ad\bd} \int d^2\si' S_F^-(\si-\si') \psi^{\bd}_{\bed}(\si'),
\eeqa
where \eg $\phi_\|(\si') \equiv \phi_\|(X_0(\si'))$. Using \Eqsref{df der}-\eqnref{sf- der} we can then rewrite the action in the following form
\beqa
S &=&  \int_\Si d^2\si \Big[ 4\pa_+\Xh'^{\al\bed} \pa_-\Xh'_{\al\bed} +
  2\eps_{\al\be}\eps^{ab} \Thh'^\al_a \pa_+\Thh'^\be_b -
  2\eps_{\ald\bed}\eps^{\ad\bd} \Thh'^{\ald}_{\ad} \pa_-\Thh'^{\bed}_{\bd} \Big]+\nn\\
&&{}+ \int_\Si d^2\si \int_\Si d^2\si' \Big[ -\frac{1}{4T} \phi^{a\bd}(\si) \de^{2}(\si-\si') \phi_{a\bd}(\si') +\nn\\
&&{}+ \frac{i}{4T}\eps^{\al\ga} \eps^{\bed\ded} \big( h_{\al\bed}(\si) - 2\pa_{\al\bed} \phi_\|(\si) \big) D_F(\si-\si') \big( h_{\ga\ded}(\si') - 2\pa_{\ga\ded} \phi_\|(\si') \big) -\nn\\
&&{}- \frac{i}{8T} \eps^{\al\be} \eps_{ab} \psi^a_\al(\si) S^+_F(\si-\si') \psi^b_\be(\si')+\nn\\
\eqnlab{almost S_eff}
&&{}+ \frac{i}{8T} \eps^{\ald\bed} \eps_{\ad\bd} \psi^{\ad}_{\ald}(\si) S^-_F(\si-\si') \psi^{\bd}_{\bed}(\si') \Big]
\eeqa
The first line in this expression contains the kinetic terms for the shifted string waves. Since the shifts of variables do not affect the measure in $Z[S]$, we find from the definitions in \Eqsref{s_eff def}-\eqnref{Z} that the effective action simply equals the last four lines of the action \eqnref{almost S_eff} above. It is not difficult to check that $S_{\sss{eff}}$ is invariant under the supersymmetry transformations \Eqsref{tm susy phi}-\eqnref{tm susy h} for the generators $\eta_a^{\ad}$ and $\eta_{\ad}^\al$ which are left unbroken by a BPS saturated string. Terms in the effective action involving a string wave propagator of course correspond to scattering processes in which string waves are participating, as described in section \secref{sub action}. The first term in $S_{\sss{eff}}$ corresponds to a particle of type $4_0$ simply bouncing of the string. It is indeed interesting to see that we have these two fundamentally different kinds of interaction processes for the tensor multiplet particles.

\subsection{The differential cross section}
\seclab{cross section}

To read off the desired scattering amplitudes, we now insert the Fourier expansions of section \secref{tm fields}. After some tedious calculations, we then end up with the following form of the effective action
\beqa
S_{\sss{eff}} &=& - \frac{(2\pi)^2}{16T} \int \frac{d^5p}{(2\pi)^{5/2}} \int \frac{d^5p'}{(2\pi)^{5/2}} \frac{1}{|\pbf|} \de(|\pbf| - |\pbf'|) \de(\pbf \cdot \nbf - \pbf' \cdot \nbf) \times\nn\\
&& \Big[ \frac{1}{2} a^*_{a\bd}(\pbf') a^{a\bd}(\pbf) + \nn\\
&&{}+ \frac{1}{\sqrt{2}}\big( a^*_{\phi_\|}(\pbf') + a^*(\pbf',0) \big) \frac{1}{\sqrt{2}}\big( a_{\phi_\|}(\pbf) + a(\pbf,0) \big) e^{i\rho} +\nn\\
&&{}+  \frac{1}{\sqrt{2}}\big( a^*_{\phi_\|}(\pbf') - a^*(\pbf',0) \big) \frac{1}{\sqrt{2}}\big( a_{\phi_\|}(\pbf) - a(\pbf,0) \big) e^{-i\rho} +\nn\\
&&{}+ a^*(\pbf',+1) a(\pbf,+1) e^{i\rho} + a^*(\pbf',-1) a(\pbf,-1) e^{-i\rho} +\nn\\
&&{}+ a^*_a(\pbf',{\sss +\frac{1}{2}}) a^a(\pbf,{\sss+\frac{1}{2}}) e^{i\rho} + a^*_a(\pbf',{\sss-\frac{1}{2}}) a^a(\pbf,{\sss-\frac{1}{2}}) e^{-i\rho} +\nn\\
&&{}+ a^*_{\ad}(\pbf',{\sss+\frac{1}{2}}) a^{\ad}(\pbf,{\sss+\frac{1}{2}}) + a^*_{\ad}(\pbf',{\sss-\frac{1}{2}}) a^{\ad}(\pbf,{\sss-\frac{1}{2}}) \Big].
\eeqa
We see that the angular dependence agrees perfectly with the scattering diamond \eqnref{scattering diamond} of section \secref{kinematics}. To get a better feeling for the angle $\rho$, note that 
\beq
p'_{\al\bed}p^{\al\bed}=\frac{1}{2}|\pbf'||\pbf| \sin^2{\theta} \cos{\rho},
\eeq
given that $\pbf'\cdot \nbf = \pbf \cdot \nbf$. This means that $\rho$ is the angle between the parts of  $\pbf'$ and $\pbf$ that are orthogonal to the string direction $\nbf$.

To obtain the $T$-matrix from this effective action we quantize the theory by means of section \secref{quantization}, so that the functions $a(\pbf)$ become operators on a Fock space. It is then a straight forward exercise to calculate the elements of the $T$-matrix by squeezing the effective action between two one-particle states. For example, the $T$-matrix for scattering between an incoming particle with polarization $2_{{\sss +\frac{1}{2}}}$ and momentum $\pbf$ and an outgoing particle with the same polarization and momentum $\pbf'$ is calculated as
\beq
\left< 0 \right| a^b(\pbf',{\sss+\frac{1}{2}}) S_{\sss{eff}} a^\dagger_{a}(\pbf,{\sss+\frac{1}{2}}) \left| 0 \right> = -\frac{|\pbf|^2}{16T(2\pi|\pbf|)^3} \de^b_a \de(|\pbf| - |\pbf'|) \de(\pbf \cdot \nbf - \pbf' \cdot \nbf) e^{i\rho}.
\eeq
The factor $-\frac{1}{16}$ has no significance, since it is an artifact of the overall constant in the free tensor multiplet action \eqnref{tm action}. (In principle, there is a correct choice of that constant. Its determination would presumably involve some kind of topological argument. At the moment, it is not clear to us exactly how to do this and therefore we have made no effort in getting it right.)

Proceeding like this, we reproduce exactly the results of section \secref{kinematics};  see especially \Eqsref{t matrix form} and  \eqnref{f}-\eqnref{scattering matrix}. Most importantly, we find that the function $f(|{\bf p}|^2 / T, \theta, \rho)$ in \Eqnref{f} is proportional to $|{\bf p}|^2 / T$, but does not depend on the angles $\theta$ and $\rho$.  We stress that these results are valid only to lowest order in perturbation theory.

We can also obtain an expression for the differential cross sections of the scattering processes in question. In five spatial dimensions, this cross section is a volume. Let us start by rewriting slightly the relation between the $T$-matrix and the invariant matrix elements in \Eqnref{t matrix form} as
\beq
T(\pbf,s \rightarrow \pbf',s') = \frac{1}{(2\pi|\pbf|)^3} \delta (|{\bf p}| - |{\bf p}^\prime|) \delta ({\bf p} \cdot {\bf n} - {\bf p}^\prime \cdot {\bf n})  \mathcal{M}(\pbf,s \rightarrow \pbf',s'),
\eeq
where, according to the above together with \Eqnref{f},
\beq
\mathcal{M}(\pbf,s \rightarrow \pbf',s') = \de_{ss'} f(|{\bf p}|^2 / T, \theta, \rho) g(\rho,s) \sim \de_{ss'} \frac{|{\bf p}|^2}{T}  g(\rho,s).
\eeq
The functions $g(\rho,s)$ are given in \eqnref{scattering diamond}. It is now fairly easy to derive the following expression for the desired differential cross sections
\beq
\eqnlab{cs}
\frac{d\si}{d\Om_3} = \frac{2}{(2\pi |\pbf|)^3} \left| \mathcal{M}(\pbf,s \rightarrow \pbf',s') \right|^2 \sim \frac{2}{(2\pi |\pbf|)^3} \frac{|\pbf|^4}{T^2}\de_{ss'}.
\eeq
Here, $d\Om_3$ is an infinitesimal element of the three-dimensional unit sphere that is orthogonal to the string. In the final step, we have used that $|g(\rho,s)|=1$ for all polarizations $s$. A priori one would have expected the differential cross section to depend on both the angles $\theta,\rho$ as well as on the polarization label $s$. However, with the particular choice of polarization basis (\ref{sbasis}) we get the very simple expression in \eqnref{cs} for all values of $s$.

\newpage

\appendix
\section{Notation and conventions}
\seclab{notation}

\subsection{$SO(5,1)$ Weyl indices}
\seclab{weyl}
We introduce Weyl spinor indices $\alh,\beh=1,...,4$. A subscript
(superscript) index denotes the (anti) Weyl representation. Single
Weyl indices cannot be raised or lowered, however antisymmetric pairs
of indices can. We then make use of the totally antisymmetric
$SO(5,1)$ invariant tensors $\eps^{\alh\beh\gah\deh}$ and
$\eps_{\alh\beh\gah\deh}$ in the following way
\beqa
\eqnlab{raising}
A^{\alh\beh} &=& \frac{1}{2} \eps^{\alh\beh\gah\deh}
A_{\gah\deh}\\
\eqnlab{lowering}
A_{\alh\beh} &=& \frac{1}{2} \eps_{\alh\beh\gah\deh}
A^{\gah\deh},
\eeqa
where $A^{\alh\beh}=-A^{\beh\alh}$ is a Lorentz vector. Note that
$\eps^{\alh\beh\gah\deh} \eps_{\alh'\beh\gah\deh} =
6\de^{\alh}_{\alh'}$.

We relate the Lorentz vector $A^{\alh\beh}$ to the familiar vector
index notation as
\beqa
A^{\alh\beh} &=& \frac{1}{2} (\Ga^\mu)^{\alh\beh} A_\mu \\
A_{\alh\beh} &=& \frac{1}{2} \epsilon_{\alh\beh\gah\deh}
A^{\gah\deh} = \frac{1}{2}(\Ga^\mu)_{\alh\beh}A_\mu,
\eeqa
where the Gamma-matrices obey the Clifford algebra
\beq
\eqnlab{clifford}
(\Ga^\mu)_{\alh\beh}(\Ga^\nu)^{\alh\gah} +
(\Ga^\nu)_{\alh\beh}(\Ga^\mu)^{\alh\gah} = 2\de^{\gah}_{\beh}
\eta^{\mu\nu}.
\eeq
As usual, $\mu,\nu=0,...,5$ and $\eta^{\mu\nu}$ is the flat Minkowski
spacetime metric with signature $(-,+,...,+)$.

These definitions imply that $\pa_{\alh\beh} x^{\alh\beh} = \pa_\mu x^\mu = 6$.

\subsection{Light-cone notation}
\seclab{l.c. conventions}
For any Lorentz vector $A^\mu$,
$\mu=0,...,5$ we write
\beqa
\eqnlab{l.c.}
A^+ &\equiv& A^0 + A^5\\
A^- &\equiv& A^0 - A^5\\
&\Rightarrow& \nn\\
A_+ &=& \frac{1}{2}(A_0+A_5)\\
A_- &=& \frac{1}{2}(A_0-A_5).
\eeqa
To relate this to the notation of $SU(2)$-indices introduced in section \secref{sub string}, we start by defining
$\eps^{\al\be}$ and $\eps^{\ald\bed}$ as
\beq
\eps^{\al\be\ald\bed} \equiv \eps^{\al\be} \eps^{\ald\bed},
\eeq
where
\beqa
\eps^{\al\be} \eps_{\ga\be} &=& \de^\al_\ga\\
\eps^{\ald\bed} \eps_{\gad\bed} &=& \de^{\ald}_{\gad}
\eeqa
and $\eps^{12}=1$ for both $\eps^{\al\be}$ and $\eps^{\ald\bed}$.

It then follows from \eqnref{lowering} that
\beqa
A_{\al\ald} = - \eps_{\al\be} \eps_{\ald\bed} A^{\be\bed}\\
\eps^{\al\be} A_{\al\be} = \eps_{\ald\bed} A^{\ald\bed}.
\eeqa
The $SU(2)$-notation is now related to the light-cone notation
by the following conventions
\beqa
\eqnlab{notation1}
A_{\al\be} &\equiv& \eps_{\al\be} A_+\\
A^{\al\be} &\equiv& \frac{1}{2} \eps^{\al\be} A^+.
\eeqa
Then, by the condition that $A^+=-2A_-$ and $A^-=-2A_+$, we find that
\beqa
\eqnlab{notation2}
A_{\ald\bed} &=& -\eps_{\ald\bed} A_-\\
A^{\ald\bed} &=& -\frac{1}{2}\eps^{\ald\bed} A^-.
\eeqa
Finally, we note that
\beq
\pa_{\alh\beh} x^{\alh\beh} = \pa_{\al\be}
x^{\al\be} + \pa_{\ald\bed} x^{\ald\bed} + 2 \pa_{\al\bed} x^{\al\bed}
= 1+1+4,
\eeq
where we have used that $\pa_{\al\ald} x^{\be\bed} =
\de^\al_\be \de^{\ald}_{\bed}/2$ in the last step.

\vspace{1cm}
\noindent
\textbf{Acknowledgments:}  I want to thank my supervisor M\aa ns Henningson for his guidance, and for substantial help in writing section \secref{kinematics} on symmetries of the scattering problem. P\"ar Arvidsson and Per Salomonson are also acknowledged for discussions.

\newpage

\bibliographystyle{utphysmod3b}
\bibliography{paper}

\end{document}